\documentclass[aps,prl,twocolumn,amsmath,amssymb,superscriptaddress,floatfix]{revtex4-1}

\usepackage{graphicx}
\usepackage{multirow}
\usepackage{color}
\usepackage{bm}
\usepackage{hyperref}

\def\t#1{\textrm{#1}}

\def\n{\nonumber \\ }

\begin{document}

\title{
Chiral anomaly and giant magnetochiral anisotropy 
in noncentrosymmetric Weyl semimetals
}

\author{Takahiro Morimoto}
\affiliation{Department of Physics, University of California, Berkeley, CA 94720, USA}
\author{Naoto Nagaosa}
\affiliation{RIKEN Center for Emergent Matter Science 
(CEMS), Wako, Saitama, 351-0198, Japan}
\affiliation{Department of Applied Physics, The University of 
Tokyo, Tokyo, 113-8656, Japan}

\date{\today}

\begin{abstract}
We theoretically propose that giant magnetochiral anisotropy is achieved in Weyl semimetals in noncentrosymmetric crystals as a consequence of the chiral anomaly. The magnetochiral anisotropy is the nonlinearity of the resistivity $\rho$ that depends on the current $\bm{I}$ 
and the magnetic field $\bm{B}$ as 
$\rho=\rho_0 (1 + \gamma \bm{I} \cdot \bm{B})$, and can be applied to rectifier devices controlled by $\bm{B}$.
We derive the formula for the coefficient $\gamma$ in noncentrosymmetric Weyl semimetals. 
The obtained formula for $\gamma$ shows that the magnetochiral anisotropy is strongly enhanced when the chemical potential is tuned to Weyl points, and that noncentrosymmetric Weyl semimetals such as TaAs can exhibit much larger magnetochiral anisotropy than that observed in other materials so far. 
\end{abstract}

\pacs{72.10.-d,73.20.-r,73.43.Cd}
\maketitle

Relativistic electronic states in solids attract recent intensive interests~\cite{Vishwanath}. 
They include two-dimensional Dirac electrons in graphene~\cite{graphene}, 
surface states of three-dimensional topological insulators~\cite{TI1,TI2}, 
band crossing in three-dimensional ferromagnetic metals 
behaving as magnetic monopoles~\cite{Fang}, 
and Weyl semimetals~\cite{Murakami,Qi}.   
In particular, the magneto-transport phenomena in Weyl semimetals 
have been actively studied in the context of chiral anomaly~\cite{Vishwanath,Qi}.
Symmetries play crucial role in the appearance of relativistic electrons.
Time-reversal symmetry $T$ and the spatial inversion symmetry $P$
are the most fundamental, in which the energy dispersion of the 
Bloch electrons satisfies the following constraints:  
$\varepsilon_{\sigma}(\bm{k}) = \varepsilon_{\bar \sigma}(-\bm{k}) $
($\bar \sigma$ means the opposite spin to $\sigma$) 
in the presence of $T$-symmetry, while $P$ symmetry imposes the
relation $\varepsilon_{\sigma}(\bm{k}) = \varepsilon_{\sigma}(-\bm{k}) $.
When both $T$ and $P$ symmetries are present,
there occurs the Kramers degeneracy at every $\bm{k}$-point and the
band crossings are described by $4 \times 4$ Hamiltonian.
In this case, additional symmetry such as point group symmetry is required to realize stable massless Dirac electrons~\cite{Yang}.
This Kramers degeneracy at each $\bm{k}$-point is lifted by the
broken $T$~\cite{Fang} or $P$ combined with the relativistic spin-orbit interaction. 
This results in the band crossing described by the 2$\times$2 Hamiltonian expanded as
$H(\bm{k}) = \sum_{\alpha = 0}^3 h_\alpha (\bm{k}) \sigma^\alpha$
near the crossing point $\bm{k}_0$ (Weyl point). Here $\sigma^0$ is the unit matrix while
$\bm{\sigma} = ( \sigma^1, \sigma^2, \sigma^3)$ are the Pauli matrices.
Expanding $h_\alpha (\bm{k})$ with respect to $\bm{k} - \bm{k}_0$ up to the linear order, one obtains the Weyl fermion (WF). After an appropriate coordinate 
transformation and neglecting $h_0(\bm{k})$ which simply gives the shift 
of the energy, one obtains  
\begin{equation}
H(\bm{k}) = \eta v_F \hbar\bm{k} \cdot \bm{\sigma}
\label{eq:WF}
\end{equation}
with the Fermi velocity $v_F$,
where $\eta = \pm 1$ determines the chirality.
Weyl semimetals are realized when the WFs are the only low-energy excitations at the Fermi energy and the transport properties are 
governed by WFs. 

 From the symmetry point of view, 
Weyl semimetals are classified according to the symmetries, i.e.,
whether $T$ or $P$ is broken.
Magnetic materials break $T$-symmetry and are able to support the Weyl semimetals.
One example is the pyrochlore antiferromagnets where the 
pairs of Weyl electrons appear along the four equivalent directions in 
momentum space~\cite{Wan11,Vishwanath}.
In this system, $P$-symmetry is intact for the single crystal 
and the WFs with opposite chiralities are located at $\bm{k}_0$ and $-\bm{k}_0$.
The other class is the noncentrosymmetric Weyl semimetals which include a recently discovered material realization of TaAs~\cite{TaAs-theory, TaAs1,TaAs2}. In the present paper, we focus on the latter noncentrosymmetric Weyl semimetals. 

The WF is characterized by the 
Berry curvature $\bm{b}(\bm{k})$ in momentum space. 
The Berry curvature $\bm{b}$ for the lower energy state for Eq.~(\ref{eq:WF}) is given by 
\begin{equation}
\bm{b}(\bm{k}) = \eta \frac{\bm{k}}{2 |\bm{k}|^3}
\label{eq:bk}
\end{equation}
which corresponds to that of monopole (anti-monopole) 
for $\eta=1$ ($\eta=-1$).
Namely, the integral of $\bm{b}(\bm{k})$ over the surface enclosing the 
Weyl point is a topological index which gives stability to the WFs.
In this case, the only way to destroy them is
the pair annihilation of two WFs with $\eta=1$ and $\eta = -1$. 
Now, the time-reversal symmetry $T$ connects the electronic states at $\bm{k}$ and 
$- \bm{k}$. As for the Berry curvature $\bm{b}(\bm{k})$, $T$ imposes the relation 
$\bm{b}(- \bm{k}) =- \bm{b}(\bm{k})$. This relation indicates that WF at $\bm{k}_0$ 
is always accompanied with its partner WF at $- \bm{k}_0$ with the {\it same} chirality.
This is because the surface integral of $\bm{b}(\bm{k})$ around $\bm{k}_0$ is the same
as that around $-\bm{k}_0$ as seen in Fig.~\ref{fig: WF}.
There is also a theorem by Nielsen-Ninomiya~\cite{Nielsen1,Nielsen2} 
that the WFs with opposite chiralities are
always paired. Therefore, it is concluded that there must be at least another pair
of WFs at  $\bm{k}_1$ and  $-\bm{k}_1$ with the opposite chirality. 
This means that in the Weyl semimetals with broken $P$ symmetry, 
there are at least four WFs, two of which have $\eta=1$ at $\pm \bm{k}_0$
and the other two have $\eta=-1$ at $\pm \bm{k}_1$. This situation is schematically 
shown in Fig.~\ref{fig: WF}.

Electromagnetic responses of WFs have unique features that originate from the 
Landau levels (LLs) formed by the magnetic field $\bm{B}$. 
In the presence of $\bm B$, so called ``zero-modes''
with the one dimensional dispersion along the direction of $\bm{B}$ are formed by the zeroth LLs. They allow electrons to be pumped with the applied electric field $\bm{E}$ 
from one Weyl node to another Weyl node of opposite chirality.
Specifically, application of both $\bm E$ and $\bm B$ in a parallel way increases the imbalance
of electron numbers $n_{\eta}$ between the WF with opposite chiralities~\cite{Nielsen3}. 
This is expressed for $\nu$ pairs of WF and anti-WFs by the equation~\cite{Burkov,Son} 
\begin{equation}
\frac{d Q^5}{d t} = \frac{2\nu}{(2\pi)^2}\frac{e^2}{\hbar^2}\bm{E} \cdot \bm{B},
\label{eq:ano}
\end{equation}
where $Q^5 = n_{\eta=1} - n_{\eta=-1}$. 
In solids, there exists the relaxation due to impurity or phonon scattering with which 
a non-equilibrium steady state is realized with 
\begin{equation}
Q^5 = \frac{\nu e^2 \tau_\t{inter}}{4\pi^2 \hbar^2} \bm{E} \cdot \bm{B},
\label{eq:steady}
\end{equation}
where $\tau_\t{inter}$ is the relaxation time of electrons for the inter-Weyl-node scattering.
This imbalance of electron numbers between WFs and anti-WFs lead to the chemical 
potential difference $\mu^5$ between different chiralities.
Chiral magnetic effect (CME) discussed in Ref.~\cite{Fukushima,Shudan}  is expressed by the 
equation for the current density $\bm{J}$ as
$\bm{J} = -(e^2/h^2) \mu^5 \bm{B}$.
This effect is derived 
from the axion electrodynamics action 
$L_\t{ax} \propto \theta(r,t) \bm E \cdot \bm B$ which is obtained from the Fujikawa Jacobian 
with $\theta(r,t)$ being the angle corresponding to the chiral gauge transformation\cite{Fujikawa}. 
Using Eq.~({\ref{eq:steady}) with this equation for CME, one obtains
$\bm{J} = -(e^4 v^3/8\pi^2 \hbar \epsilon^2)\tau_\t{inter} (\bm{E} \cdot \bm{B}) \bm{B}$, 
where $\epsilon$ is the chemical potential measured from the Weyl point. 
This describes the magneto-transport i.e., the linear response to 
the electric field $\bm{E}$ that is modified by the external magnetic field. 
Such current response $J \propto B^2 E$ is allowed in both $T$-broken and $P$-broken Weyl semimetals \cite{Fukushima}. 
In a similar manner, we can consider another current response $J \propto B E^2$ in the case of $P$-broken Weyl semimetals as we discuss below.

\begin{figure}
\begin{center}
\includegraphics[width=0.95\linewidth]{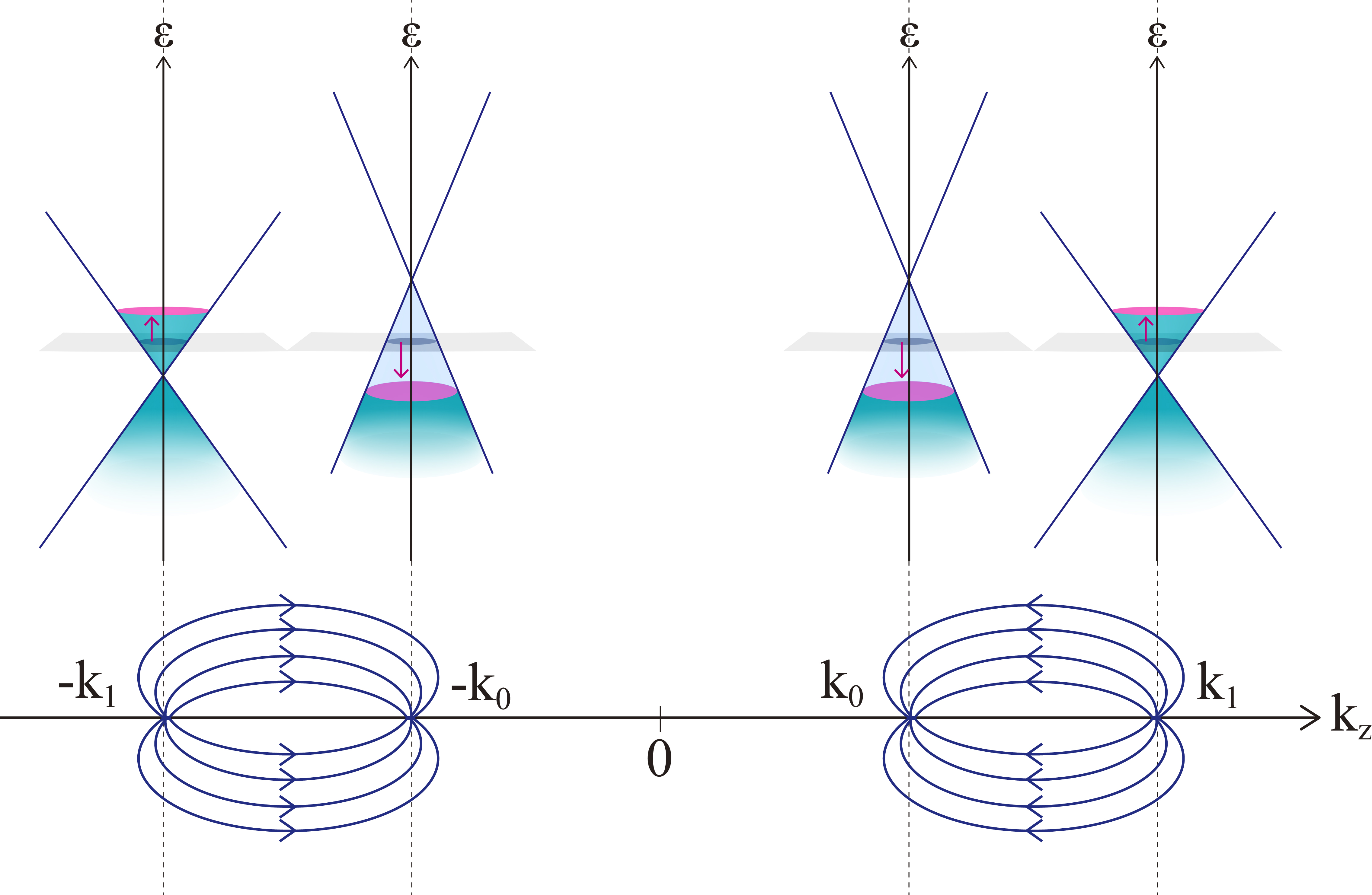}
\caption{\label{fig: WF}
Schematic picture of Weyl fermions in noncentrosymmetric system.
Time-reversal symmetry connects the flow of the Berry curvature
$\bm{b}(\bm{k})$ at $\bm{k}$ to that at $-\bm{k}$ 
as $\bm{b}(-\bm{k})= -\bm{b}(\bm{k})$. This means that the 
Weyl fermions at $\bm{k}$ and $- \bm{k}$ has the same chirality, 
i.e., both are monopoles or anti-monopoles. Therefore, there must be at least
4 Weyl fermions in noncentrosymmetric Weyl semimetals with time-reversal symmetry,
i.e., two pairs of WFs and anti-WFs, respectively.
When both the magnetic ($\bm{B}$) and electric ($\bm{E}$) fields are applied, 
the charge transfer between Weyl points occurs between monopoles 
and anti-monopoles due to chiral anomaly. This phenomenon is shown by the shift of the chemical potentials (arrows) in the figure, which drives the system into a nonequilibrium state.
Anisotropy between WFs and anti-WFs (which is allowed by the broken inversion symmetry) leads to nonreciprocal current response in this nonequilibrium state induced by the chiral anomaly. 
}
\end{center}
\end{figure}

One of the interesting effects in noncentrosymmetric systems is the
nonreciprocal response, i.e., the propagation of light or the flow of 
current that depends on the direction, with the external magnetic field $\bm{B}$ or the
spontaneous magnetization $\bm{M}$ that breaks 
$T$-symmetry~\cite{RikkenNature,RikkenBi,RikkenMol,RikkenCNT,RikkenSi}.
The time-reversal symmetry in the microscopic 
dynamics leads to the Onsager's reciprocal relation, which 
imposes a condition on the conductivity tensor 
$\sigma_{ij}$ as
\begin{align}
\sigma_{ij}(\bm{k}, \bm{B}) = \sigma_{ji}(- \bm{k}, -\bm{B}),
\label{eq:Onsager}
\end{align}
with the wavevector $\bm{k}$,
and governs the form of the nonreciprocal responses. 
A well known example of nonreciprocal responses is the optical magnetochiral dichroism 
that is realized when
the magnetic field $\bm{B}$ (or magnetization $\bm{M}$) 
and the electric polarization $\bm{P}$ 
form the toroidal moment $\bm{T} = \bm{P} \times \bm{B}$ (or
$\bm{T} = \bm{P} \times \bm{M}$)~\cite{RikkenNature}.
In this case, the dielectric constant $\varepsilon$ 
depends on the relative direction of the pointing vector $\bm{S}$ of the light ($\propto \bm k$)
and the toroidal moment $\bm{T}$.
Phenomenologically, this can be expressed as
$\varepsilon = \varepsilon_0 + \alpha \bm{T} \cdot \bm{k}$ where 
$\bm{k}$ is the wavevector of light. In fact, this nonreciprocal linear response is consistent with the Onsager's relation in Eq.~(\ref{eq:Onsager}). 
Meanwhile, nonreciprocal responses in the transport phenomena 
have been studied by Rikken et al.~\cite{RikkenBi}.
They discussed that the the wavevector $\bm{k}$ can be replaced by the
velocity or current of the electrons $\bm{I}$ in Eq.(\ref{eq:Onsager}) and, hence, the $\bm I \cdot \bm B$ term is allowed in the conductivity tensor,
which leads to the transport magnetochiral 
anisotropy~\cite{RikkenBi,RikkenMol,RikkenCNT}.
Specifically, the transport magnetochiral 
anisotropy is the current response $J \propto B E^2$ and is described by the resistivity $\rho$ that depends on the current 
$\bm{I}$ and the external magnetic field $\bm{B}$ as 
\begin{equation} 
\rho = \rho_0( 1 + \gamma \bm{I} \cdot \bm{B}).
\label{eq:CM}
\end{equation}
This effect has been studied for Bi helix~\cite{RikkenBi}, molecular solid~\cite{RikkenMol},
 and carbon nanotubes~\cite{RikkenCNT}
with chiral structure,
where two major microscopic mechanisms have been proposed. 
One is the magnetic self-field.
In the presence of the magnetic self-field,
the magnetoresistance is expressed by 
$\Delta \rho = \beta \bm{B}_\t{eff}^2$,
where $\bm{B}_\t{eff}^2$ is the effective magnetic field given by the sum of
the external one $\bm{B}_{\rm ext}$ and $\bm{B}'$ induced by the current $\bm{I}$. Due to the helical structure of the materials,
the induced $\bm{B}'$ is parallel to $\bm I$ and produces positive magnetoresistance resulting in Eq.(\ref{eq:CM}). 
The other mechanism is the scattering of electrons by chiral objects
such as crystal defects and phonons.
It has been also found that the spin-orbit interaction in Si leads to a different 
type of the magnetochiral anisotropy of the form
$\rho = \rho_0 [ 1 + \chi \bm{E} \cdot (\bm{I} \times \bm{B})]$
where the external electric field $\bm{E}$ plays a role of the inversion symmetry breaking~\cite{RikkenSi}.
The strength of the magnetochiral anisotropy is usually discussed in terms of the
coefficient $\gamma$.
However, $\gamma$ depends on the cross section $A$ of the sample, and more intrinsic quantity is $\gamma'=\gamma A.$
The coefficients $\gamma$ and $\gamma'$ have been measured experimentally as
$\gamma \simeq 10^{-3}$T$^{-1}$A$^{-1}$ and $\gamma' \simeq 10^{-10}$m$^2$T$^{-1}$A$^{-1}$ in Bi helix~\cite{RikkenBi},
$\gamma \simeq 10^{-3}$T$^{-1}$A$^{-1}$ and $\gamma' \simeq 10^{-11}$m$^2$T$^{-1}$A$^{-1}$ in molecular solid~\cite{RikkenMol},
$\gamma \simeq 10^{-1}$T$^{-1}$A$^{-1}$ and $\gamma' \simeq 10^{-10}$m$^2$T$^{-1}$A$^{-1}$ in Si~\cite{RikkenSi} \footnote{Although the cross section of the Si sample is not explicitly written in Ref.~\cite{RikkenSi}, we assumed the cross section to be $A=10^{-9}\t{m}^2$ by considering the channel length of $100\mu$m.},
and 
$\gamma \simeq 10^{2}$T$^{-1}$A$^{-1}$ and $\gamma' \simeq 10^{-16}$m$^2$T$^{-1}$A$^{-1}$ in carbon nanotube~\cite{RikkenCNT}. 
From the viewpoint of the applications of magnetochiral anisotropy as rectifying 
function, the larger values of $\gamma$ and $\gamma'$ are desirable
because it enables more efficient rectifier devices which are controllable with magnetic fields.

Since the Weyl semimetal has been realized in noncentrosymmetric materials such as TaAs, it is an interesting issue to study Weyl semimetals as a platform for the magnetochiral anisotropy.
In particular, it is interesting to explore its relationship to the chiral anomaly which is an origin of various anomalous transport properties.
Motivated by this, we now proceed to the prediction of the magnetochiral anisotropy 
in noncentrosymmetric Weyl semimetals.
It turns out that that the inversion symmetry breaking and the chiral anomaly play a crucial role in the magnetochiral anisotropy of the Weyl semimetals as follows.
In the noncentrosymmetric Weyl semimetals, the WFs with different chiralities are 
not equivalent and usually forms the small electron or hole pockets. 
In the presence of  $\bm{B} \cdot \bm{E}$, 
the chiral anomaly triggers changes of the sizes of these inequivalent pockets, and hence, 
the value of metallic conductivity. This results in nonlinear resistivity proportional 
to $B$ in noncentrosymmetric Weyl semimetals.

To substantiate the above idea, we consider WFs at $\pm \bm {k_+}$ and anti-WFs at $\pm \bm {k_-}$ 
for which the Fermi velocity is $v_\pm$ and the Fermi energy measured 
from the Weyl point is $\epsilon_\pm$, respectively, as schematically illustrated in Fig.~\ref{fig: WF}.
The parameters $v_\pm$ and $\epsilon_\pm$ can differ for the WFs and anti-WFs due to the broken inversion symmetry.
Contributions to the linear conductivity from the WFs/anti-WFs are written as
\begin{align}
\sigma_\pm=\frac{1}{3}e^2 v_\pm^2 \tau_\t{intra} D_\pm(\epsilon_\pm).
\end{align}
Here, $\tau_\t{intra}$ is the relaxation time for intra-Weyl-node scattering which is usually shorter than $\tau_\t{inter}$~\cite{Son}, and the density of states $D_\pm(\epsilon)$ is given by
\begin{align}
D_\pm(\epsilon)=\frac{\nu \epsilon^2}{2\pi^2 \hbar^3 v_\pm^3},
\end{align}
where $\nu$ denotes the number of pairs of WFs/anti-WFs. For example, $\nu=12$ for TaAs.
This reduces to
\begin{align}
\sigma_\pm=\frac{\nu e^2}{6\pi^2 \hbar^3} \frac{\tau_\t{intra} \epsilon_\pm^2}{v_\pm}.
\label{eq: sigma pm}
\end{align}
When the electric field $\bm E$ and the magnetic field $\bm B$ are applied to the sample in a parallel way, electrons are transfered from the WFs to the anti-WFs (or vice versa) due to the chiral anomaly as shown in Fig.~\ref{fig: WF}. 
This results in changes of Fermi energies for the WF and the anti-WFs given by $\Delta \epsilon_\pm=\pm Q^5/D_\pm(\epsilon_\pm)$. 
As a result, the linear conductivity is also modified as
\begin{align}
\Delta \sigma_\pm&=\frac{1}{3}e^2 v_\pm^2 \tau_\t{intra} \frac{d D_\pm(\epsilon)}{ d\epsilon} 
\Delta \epsilon_\pm,
\end{align}
which can be explicitly written as
\begin{align}
\Delta \sigma_\pm
&= 
\pm \nu \frac{e^4}{6\pi^2 \hbar^2} \frac{v_\pm^2 \tau_\t{intra} \tau_\t{inter}}{\epsilon_\pm} 
\bm{E} \cdot \bm{B}.
\end{align}
These changes of the linear conductivity can become nonvanishing after summing over the WFs and the anti-WFs due to the anisotropy in $v_\pm$ and $\epsilon_\pm$.
In this case, the nonlinear current response $\bm J \propto (\bm E \cdot \bm B) \bm E$ is realized and supports the nonreciprocal current response; in the presence of the magnetic field $\bm B$, different magnitude of dc current flows in the case where $\bm E$ and $\bm B$ are parallel compared to the case where $\bm E$ and $\bm B$ are antiparallel. Thus the noncentrosymmetric Weyl semimetals support the current rectification effect originating from the chiral anomaly.

Now we quantify the nonreciprocal current response in Weyl semimetals  by defining the intrinsic nonlinear resistivity coefficient 
$\gamma'$ as $\gamma'=\gamma A$ with the cross section of the sample $A$.
This coefficient $\gamma'$ does not depend on the cross section of the sample and can be obtained from the conductivity change in the above as
\begin{align}
\gamma' &= -\frac{2\Delta \sigma/(\bm E \cdot \bm B)}{\sigma^2}
\n
&=
-\frac{12 \pi^2 \hbar^4 \tau_\t{inter}}{\nu \tau_\t{intra}} \left(\frac{v_+^2}{\epsilon_+}-\frac{v_-^2}{\epsilon_-} 
\right)\left(\frac{\epsilon_+^2}{v_+}+\frac{\epsilon_-^2}{v_-} \right)^{-2} 
.
\end{align}
Let us see that the expression for $\gamma'$ is simplified in the following two cases:
(i) In the case of $v_\pm=v_0$, the nonlinear resistivity coefficient $\gamma'$ further reduces to
\begin{align}
\gamma'
=
-\frac{12 \pi^2 \hbar^4 v_0^4 \tau_\t{inter}}{\nu \tau_\t{intra}} 
\frac{- \epsilon_+ + \epsilon_-}{\epsilon_+ \epsilon_- (\epsilon_+^2 + \epsilon_-^2)^2} ,
\end{align}
which is proportional to the energy difference of Weyl points ($\epsilon_+ - \epsilon_-$).
(ii) In the case of $\epsilon_+=\epsilon_-=\epsilon$, this reduces to
\begin{align}
\gamma'
=
-\frac{12 \pi^2 \hbar^4 \tau_\t{inter}}{\nu \tau_\t{intra}} \frac{(v_+ v_-)^2 (v_+ - v_-)}{(v_+ + v_-) \epsilon^5} ,
\label{eq: gamma' same epsilon}
\end{align}
which is proportional to the Fermi velocity difference ($v_+ - v_-$).
Therefore any asymmetry of band structures between WFs and anti-WFs can 
support the rectification effect proportional to $B$ originating from the chiral anomaly.
In both cases, the magnetochiral anisotropy is enhanced when the Weyl points are close to the Fermi energy and the Fermi velocity is large.
In particular, we notice by comparing Eq.~(\ref{eq: gamma' same epsilon}) and Eq.~(\ref{eq: sigma pm}) that the enhancement of magnetochiral anisotropy $\gamma'\propto \epsilon^{-5}$ with $\epsilon \to 0$ is larger than that of magnetoresistance $1/\sigma \propto \epsilon^{-2}$.

\textit{Discussions ---}
We give a crude estimate for the nonlinear ratio $\gamma$.
In time-reversal symmetric Weyl semimetals such as TaAs \cite{TaAs-theory,TaAs1}, typical parameters are given by $v=4 \times 10^5 \t{m/s}$ 
and $|\epsilon_\pm| \sim 10 \t{meV}$, and we assume $\tau_\t{intra}\simeq\tau_\t{inter}$ for simplicity. 
In this case, the estimate is given by 
$\gamma' \simeq 3 \times 10^{-8} \times \t{m}^2 \t{T}^{-1} \t{A}^{-1} $.
This coefficient $\gamma'$ for Weyl semimetals is larger than that for any of the materials reported in Refs.~\cite{RikkenBi,RikkenMol,RikkenNature,RikkenCNT}.
If we consider the sample of a cross section $A=0.1\t{mm}^2$, 
we obtain the nonlinear coefficient $\gamma=0.3~ T^{-1} A^{-1}$.
From the practical point of view, the dimensionless factor $\eta= \gamma \bm I \cdot \bm B =\gamma' \bm J \cdot \bm B$ determines the ratio between the currents of right and left directions. This rectification efficiency $\eta$ can be of the order of unity in our case for $J=10^3 \t{A}/\t{mm}^2$ and $B=0.1\t{T}$, while it is typically of the order of $10^{-4}$ - $10^{-3}$ in the systems in Refs.~\cite{RikkenBi,RikkenMol,RikkenNature,RikkenCNT}.
In this regard, Weyl semimetals which are metals and have large $\gamma'$ offer an efficient nonreciprocal property in the magnetic field.

Comments are in order for the validity of our semiclassical approach. We used the semiclassical formula Eq.~(\ref{eq: sigma pm}) for the conductivity change at each Weyl node in deriving the magnetochiral anisotropy in Weyl semimetals. 
This description is valid when the Landau level separation $\Delta \epsilon_\t{LL}$ is smaller than the level broadening $h/\tau_\t{intra}$.
When the Landau level separation becomes larger than the level broadening with a strong magnetic field $\bm B$,
the energy bands decouple into separate 1D channels of LLs along the direction of $\bm B$. In this case, the system does not exhibit the magnetochiral anisotropy by applying $\bm E$ because pumping of electrons between left movers and right movers at Weyl/anti-Weyl nodes does not induce change of the conductivity (the conductance is always $2e^2/h$ for each 1D channel). Thus the nonvanishing magnetochiral anisotropy requires that the system is in the semiclassical regime $\Delta \epsilon_\t{LL}< \hbar/\tau_\t{intra}$, which constrains the strength of magnetic field $\bm B$. Since
the Landau level separation is given by $\Delta \epsilon_\t{LL} \cong 10 \t{meV}(B/1\t{T})$ for the parameters in the above and $h/\tau_\t{intra}=4 \t{meV}$ for the relaxation time $\tau_\t{intra}=1\t{ps}$,
the magnetic field of $B<0.4 \t{T}$ justifies the semiclassical approach  and supports the large magnetochiral anisotropy in Weyl semimetals. 
In addition, Weyl fermions realized in materials such as TaAs is not isotropic in the momentum space and shows directional anisotropy (i.e., the Fermi velocity $v_i$ differs for directions $i=x,y,z$) \cite{Lee}. This modifies the formula for magnetochiral anisotropy, but the qualitative behavior such as the scaling law with $\epsilon$ remains unchanged~%
\footnote{
In the case of anisotropic Weyl fermions with the velocity $v_{\pm,i}$ for Weyl nodes with chirality $\pm$ and the direction $i=x,y,z$, the formula for the magnetochiral anisotropy is given as follows.
We consider the case of $\bm E \parallel \bm B \parallel \hat{z}$.
In this case, the conductivity along the $z$-direction is given by
$
\sigma_{\pm}=\frac{1}{3}e^2 v_{\pm,z}^2 \tau_\t{intra} D_\pm(\epsilon_\pm), 
$
with 
$D_\pm(\epsilon)=\nu\epsilon^2/2\pi^2\hbar^3 v_{\pm,x} v_{\pm,y} v_{\pm,z}$.
Therefore, the coefficient $\gamma'$ for the magnetochiral anisotropy  is written as
$$
\gamma' =
-\frac{12 \pi^2 \hbar^4 \tau_\t{inter}}{\nu \tau_\t{intra}} \left(\frac{v_{+,z}^2}{\epsilon_+}-\frac{v_{-,z}^2}{\epsilon_-} 
\right)\left(\frac{\epsilon_+^2 v_{+,z}}{v_{+,x} v_{+,y}}+\frac{\epsilon_-^2 v_{-,z}}{v_{-,x} v_{-,y}} \right)^{-2} 
.
$$
}.
Finally, we note that the effect of electron-electron interaction and disorder is captured by the relaxation times $\tau_\t{intra}$ and $\tau_\t{inter}$ in the semiclassics. This indicates that the types of the interaction or disorder affect the ratio $\tau_\t{inter}/\tau_\t{intra}$ that enters in the formula for $\gamma'$. Namely, when the long-range (short-range) interaction or long-range (short-range) scatters are dominant, $\tau_\t{inter}/\tau_\t{intra}$ becomes large (small) and enhances (suppresses) the magnetochiral anisotropy.
Moreover, the magnetochiral anisotropy ($\gamma$ and $\gamma'$) becomes even larger than the estimate above due to the factor of
$\tau_\t{inter}/\tau_\t{intra}$ because the internode relaxation time $\tau_\t{inter}$ is usually larger than intranode relaxation time $\tau_\t{intra}$.

In general there exist other rectification effects in the presence of the inversion symmetry 
breaking and time-reversal breaking. What is special for the rectification effect proposed 
here for Weyl semimetals is that it originates from the chiral anomaly of WFs and the 
direction of the rectification can be controlled by $\bm B$.
Namely, although the expression is the same as Eq.~(\ref{eq:CM}),
the crystal structure determines the direction of the current in the cases
of helix ~\cite{RikkenBi,RikkenCNT} or molecular solid~\cite{RikkenMol}.
In contrast, in the present case, the rectification effect is essentially free from the crystal anisotropy,
i.e., determined solely by the relative angle between 
$\bm{I}$ and $\bm{B}$.   
Actually, this is the signature of the negative magnetoresistance due to the
chiral anomaly~\cite{Kim,TaAs1,Ong}. Since the present magnetochiral anisotropy 
is the twin effect of this negative magnetoresistance, it is quite natural that the 
magneto chiral anisotropy shows the similar
angle dependence.
In addition, it is interesting to note that Fermi pockets of the Weyl and anti-Weyl nodes are found to be connected in TaP~\cite{Arnold}. Even in this case, we can expect the magnetochiral anisotropy since charge transfer is induced  between different parts of the Fermi surface by applying both $\bm E$ and $\bm B$ fields as a remnant of the chiral anomaly, which results in the negative magnetoresistance observed in Ref.~\cite{Arnold}. Since TaP breaks inversion symmetry, this charge transfer can also lead to change of the linear conductivity and hence the magnetochiral anisotropy. However, the isotropic form of the coefficient with $\bm I \cdot \bm B$ term is modified to some anisotropic form reflecting details of materials in this case.

To summarize, we have theoretically proposed the magnetochiral anisotropy of 
topological origin, i.e., chiral anomaly of Weyl fermion, in noncentrosymmetric 
Weyl semimetals. This effect is missing in the centrosymmetric Weyl semimetals 
with magnetism, since it is prohibited by the inversion symmetry. The magnitude of this
effect can be very large $\delta \rho/\rho_0 \sim 1$, and the peculiar angle 
dependence will be the signature of this effect as in the case of 
negative magnetoresistance. This effect may be utilized in rectifier devices 
controlled by the external magnetic field.  

\textit{Acknowledgment ---}
We thank M. Ueda, T. Hayata, M. Ezawa for fruitful discussions.
This work was supported by 
the EPiQS initiative of the Gordon and Betty Moore Foundation (TM)
and by JSPS Grant-in-Aid for Scientific Research
(No. 24224009, and No. 26103006) from MEXT, Japan, and ImPACT Program
of Council for Science, Technology and Innovation (Cabinet
office, Government of Japan) (NN).

\bibliography{Rect.bib}
\end{document}